\documentclass[10pt,twocolumn,letterpaper]{article}

\usepackage[paper=a4paper,headheight=0pt,left=1.8cm,top=1.85cm,right=1.5cm,bottom=2.3cm]{geometry}

\usepackage{times}
\usepackage{epsfig}
\usepackage{graphicx}
\usepackage{amsmath}
\usepackage{amssymb}

\usepackage[margin=10pt,font=small,labelfont=bf]{caption}

\usepackage{booktabs}
\usepackage{multirow}
\usepackage{subfigure}
\usepackage{epsfig}
\usepackage{graphicx}
\usepackage{amsmath}
\usepackage{bm, eucal,xspace}

\def\cL{{L}}

\def\bx{\bm{x}}

\def\ba{\bm{a}}
\def\bb{\bm{b}}
\def\bv{\bm{v}}
\def\bw{\bm{w}}
\def\bV{\bm{V}}
\newcommand{\etal}{\textit{et al.}\xspace}
\newcommand{\eg}{\textit{e.g.}}
\newcommand{\ie}{\textit{i.e.}}

\begin{document}

\title{Fully Quantized Image Super-Resolution Networks}

\author{Hu Wang \ \ \ \ \ \ Peng Chen \ \ \ \ \ \ Bohan Zhuang \ \ \ \ \ \ Chunhua Shen\\
[0.12cm]
{The University of Adelaide, Australia}
}

\date{}

\maketitle

\begin{abstract}
\it 
With the rising popularity of intelligent mobile devices, it is of great practical significance to develop accurate, real-time and energy-efficient image Super-Resolution (SR) inference methods. A prevailing method for improving the inference efficiency is model quantization, which allows for replacing the expensive floating-point operations with efficient bitwise arithmetic. To date, it is still challenging for quantized SR frameworks to deliver a feasible accuracy-efficiency trade-off. Here, we propose a Fully Quantized image Super-Resolution framework (FQSR) to jointly optimize efficiency and accuracy. In particular, we target obtaining end-to-end quantized models for all layers, especially including skip connections, which was rarely addressed in the literature of SR quantization. We further identify obstacles faced by low-bit SR networks and propose a novel method to counteract them accordingly. The difficulties are caused by 1) for SR task, due to the existence of skip connections, high-resolution feature maps would occupy a huge amount of memory spaces; 2) activation and weight distributions being vastly distinctive in different layers; 3) the inaccurate approximation of the quantization. We apply our quantization scheme on multiple mainstream super-resolution architectures, including SRResNet \cite{ledig2017photo}, SRGAN \cite{ledig2017photo} and EDSR \cite{lim2017enhanced}. Experimental results show that our FQSR with low-bits quantization is able to achieve on par performance compared with the full-precision counterparts on five benchmark datasets and surpass the state-of-the-art quantized SR methods with significantly reduced computational cost and memory consumption.
\end{abstract}

\section{Introduction}

The rapid development of Deep Convolutional Neural Networks (CNNs) has led to significant breakthroughs in image super-resolution, which aims to generate high-resolution images from low-resolution inputs. For real-world applications, the inference of SR is usually executed on edge devices, such as High-Definition televisions, mobile phones or drones, which require real-time, low-power consumption. 
However, the high computational cost of CNNs prohibits the deployment of SR models to resource-constrained devices.

To improve computation and memory efficiency, various solutions have been proposed in the literature, including network pruning \cite{zhuang2018discrimination,liu2018rethinking}, low-rank decomposition \cite{zhang2016accelerating}, network quantization \cite{zhuang2018towards,Liu_2018_ECCV} and efficient architecture design \cite{howard2017mobilenets,zhang2017shufflenet,liu2019darts}. In this work, we aim to train a low-precision SR network, including \textit{all layers and skip connections}. Although current quantization methods have achieved promising performance on the image classification task, training quantized models for pixel-wise dense prediction tasks such as super-resolution, still remains a challenge in terms of the unverifiable efficiency improvement on hardware and the severe accuracy degradation. For example, to our knowledge, existing quantized SR models typically keep the skip connections to be full-precision, resulting in huge memory consumption occupied by high-resolution feature maps and making it impractical to be deployed. In this paper, we introduce the Fully Quantized Image Super-Resolution Networks (FQSR), to yield a promising efficiency-versus-accuracy trade-off.

Typically, a common SR network consists of a feature extraction module, a nonlinear mapping module and an image reconstruction module
\cite{xin2020binarized}. Recently, various quantized SR methods \cite{ma2019efficient,xin2020binarized} leverage binary quantization on the non-linear mapping module of the SR network, while paying less attention to the quantization of the feature extraction and image reconstruction modules. However, we observe that the feature extraction and reconstruction modules also account for significant computational cost during inference (\eg, with respect to $\times 2$ up-scaling models, these two sub-modules occupy 15.6\% of total computational FLOPs for SRResNet model and 11.4\% for the EDSR model; in $\times 4$ up-scaling, these two sub-modules occupy 45.1\% and 38.7\% of total computational FLOPs for SRResNet and EDSR, respectively). Therefore, it is essential 
to quantize all three sub-modules to obtain more compact models. Additionally, in the SR task, the feature dimensions are usually very high. These features will occupy a huge amount of memories, especially when skip connections exist in the network which requires multiple copies of the tensors. Thus, with the quantization of the skip connections, the memory consumption can be saved dramatically (by approximately $8\times$ when compared to the full-precision counterparts). In this paper, we propose to quantize all layers in SR networks to reduce the burden of computation and storage starving SR tasks on resource-limited platforms.

In addition to the fully quantized design, we further introduce specific modifications with respect to the quantization algorithm for super-resolution. In particular, we empirically observe that the data distributions of the activations and weights of different layers differ drastically for the SR task. For the distribution with a small value range, the corresponding quantization interval should be sufficiently compact in order to maintain appropriate quantization resolution. On the other hand, if the quantization interval is too compact for distribution with a large value range, it may cause severe information loss. Therefore, we propose to learn quantizers that can find the optimal quantization intervals that minimize the task loss. 
Furthermore, we also observe that the categorical distribution of quantized values may not fit the original distribution in some layers during training. Thus, we propose a quantization-aware calibration loss to encourage the minimization of the distribution difference.

Our contributions are summarized as follows.

\begin{itemize}

\item We introduce fully quantized neural networks for image super-resolution to thoroughly quantize the model including all layers within three sub-modules. To the best of our knowledge, \textit{we are the first to perform fully end-to-end quantization for the SR task.}

\item We identify several difficulties faced by current low bitwidth SR networks. Specifically, we quantize all skip connections to tackle a huge amount of memory consumption issue caused by high-resolution feature maps. We propose quantizers with learnable intervals to adapt the vastly distinct distributions of weights and activations in different network layers. To further reduce the quantization error, we also introduce a self-supervised calibration loss to predict the categorical distribution towards the original continuous distribution.

\item Our extensive experiments with various bit configurations demonstrate that our FQSR is able to achieve comparable performance with the full-precision counterparts, while saving considerable amount of computation and memory usage.  Moreover, experiments on mainstream architectures and datasets demonstrate the superior performance of the proposed FQSR over a few competitive state-of-the-art methods.
\end{itemize}

\section{Related Work}

\noindent\textbf{Image super-resolution.} Super-resolution research has attracted increasing attention in recent years. Since the deep learning based super-resolution is first proposed by Dong \etal \cite{dong2014learning,dong2015image}, a variety of convolutional neural models have been studied. 
ESPCN \cite{shi2016real} is proposed to optimize the SR model by learning sub-pixel convolutional filters. 
Ledig \etal \cite{ledig2017photo} introduce a Generative Adversarial Networks (GANs) SR model named SRGAN, along with which the generator is described as SRResNet. 
Lim \etal \cite{lim2017enhanced} propose a model named EDSR.
Residual channel attention is introduced by Zhang \etal \cite{zhang2018image} to overcome gradient vanishing problem in very deep SR networks.

Besides, much effort has been devoted to improve the efficiency of the SR models by designing light-weight structures. 
For example, works in \cite{dong2016accelerating,shi2016real} speed up the SR without the upsampling operations. 
Hui \etal \cite{hui2019lightweight} introduce a light-weighted information multi-distillation block into the proposed super-resolution model. 

\noindent\textbf{Model quantization.}
Model quantization aims to represent the weights, activations and even gradients in low-precision, to yield highly compact DNNs. Notably, convolutions and matrix multiplications can be replaced with bitwise operations, which can be implemented more efficiently than the floating-point counterpart. In general, quantization methods involve binary neural networks (BNNs) and bitwise quantization. In particular, BNNs \cite{rastegari2016xnor,hubara2016binarized,zhuang2019structured,
Liu_2018_ECCV} constrain both weights and activations to only two possible values (\eg, $+1$ or $-1$), enabling the multiply-accumulations being replaced by the bitwise operations: $\rm{xnor}$ and $\rm{bitcount}$. However, BNNs usually suffer from severe accuracy degradation. To make a trade-off between accuracy and efficiency, researchers also study bitwise quantization with higher-bit representation. To date, most quantization techniques employ uniform quantizers to fit the data, based on statistics of the data distribution \cite{zhou2016dorefa,Cai_2017_CVPR}, minimizing quantization error during training \cite{choi2018pact,zhang2018lq} or minimizing the task loss with stochastic gradient descent \cite{jung2019learning,esser2019learned,zhuang2018towards}.

In terms of quantization for super-resolution, Ma \etal \cite{ma2019efficient} apply BNNs to compress super-resolution networks. 
Note that it only proposes to binarize the weights of the residual blocks within the model. Most recently, Xin \etal \cite{xin2020binarized} propose a bit-accumulation mechanism for single image super-resolution to boost the quantization performance. 
Both the weights and activations are quantized, however, their models are only partially quantized, with the feature extraction module, image reconstruction module and skip connections keeping in full precision. In contrast, our FQSR 
quantize all layers within three sub-modules and skip connections, which delivers improved efficiency and accuracy trade-off.

\section{Method}
\subsection{Preliminary}

In this work, we propose to quantize weights of all convolutional layers and activations of all the network layers into low-precision values. According to \cite{rastegari2016xnor,zhou2016dorefa}, 
for two binary vector $\ba \in \{0,1\}^N$ and $\bb \in \{0,1\}^N$ within binary neural networks (BNNs), the inner product of them can be formulated as:

\begin{equation}\label{equ:bnn_dot}
    \ba \cdot \bb= {\rm{bitcount}}(\ba \operatorname{\&} \bb),
\end{equation}
where $\rm{bitcount}$ counts the number of bits in a bit vector and $\&$ represents the bitwise ``$\rm{and}$'' operation.

More generally, for quantization with higher and arbitrary bit-widths, the quantized values can be viewed as the linear combination of binary bases. Let $\ba$ be a $M$-bit quantized vector which can be represented as $\ba =\sum_{m=0}^{m=M-1}{\ba}_m\cdot 2^m$, where ${\ba}_m \in \{0, 1\}^N$. Similarly, for another $P$-bit vector $\bb$, we have $\bb =\sum_{p=0}^{p=P-1}{\bb}_p\cdot 2^p$, where ${\bb}_p \in \{0, 1\}^N$. Formally, the inner product calculation between $\ba$ and $\bb$ is
\begin{equation}
    \ba \cdot \bb=\sum_{m=0}^{M-1} \sum_{p=0}^{P-1} 2^{m+p} {\rm{bitcount}} \left(\ba_m \operatorname{\&} \bb_p\right).
\end{equation}

For a general full-precision value $v$ (activation or weight) to be quantized, an interval parameter is introduced to control the quantization range. 
The quantization function can be formulated as:
\begin{equation}\label{equ:quant}
    Q(v)=\left\lfloor\operatorname{clip}\left(\frac{v}{I}, -Q_{\rm{min}},Q_{\rm{max}} \right) \times (2^M-1)\right\rceil \times \frac{I}{2^M-1},
\end{equation}
where $I$ represents the quantization interval, $2^M$ presents the quantization levels for $M$-bit quantization, $\lfloor v \rceil$ rounds $v$ to the nearest integer, and $\operatorname{clip}\left(v, v_\mathrm{low}, v_\mathrm{up}\right) = {\rm min}[{\rm max}(v,v_\mathrm{low}),v_\mathrm{up}]$.
For unsigned data, $Q_{\rm min}=0$ and $Q_{\rm max}=1$; for signed data, $Q_{\rm min}=-1$ and $Q_{\rm max}=1$. 
At the end of the equation, a scale factor $\frac{I}{2^M-1}$ is multiplied to the intermediate results after rounding operation to re-scale the value back to its original magnitude. In our paper, practically, we privatize quantizers for activations and weights in each layer.

During the training process, latent full-precision weights are kept to update the gradients during back-propagation, while being discarded during inference. 
The gradient is derived by using the straight-through estimator (STE) \cite{bengio2013estimating} to approximate the gradient through the non-differentiable rounding function as a pass-through operation, and differentiating all other
operations in Eq.~(\ref{equ:quant}) normally.

\begin{figure}[htb]
\centering
\scalebox{0.52}{
\centerline
{
\includegraphics[width=1\textwidth]{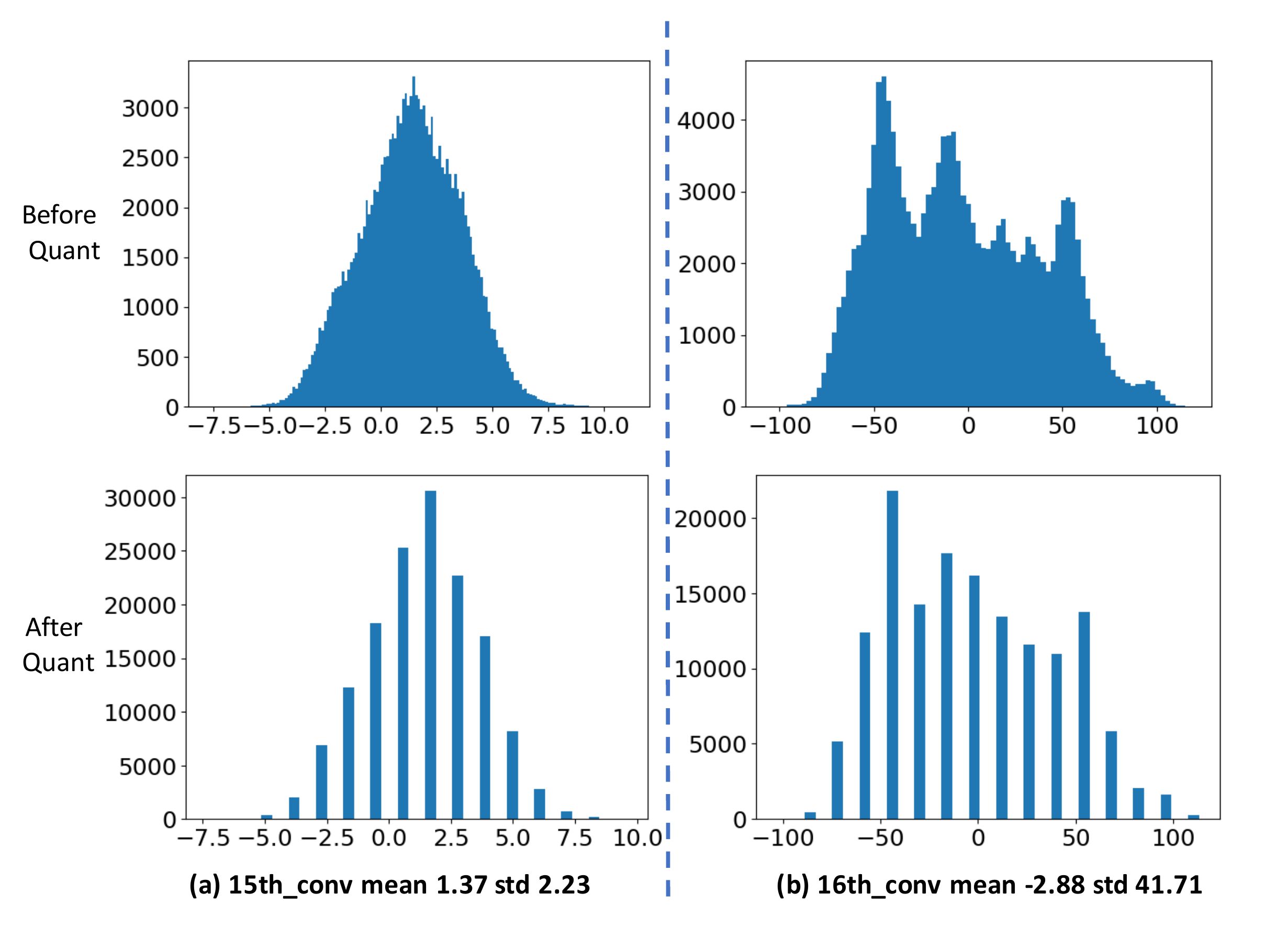}
}
}
\caption{In the figure, (a) and (b) are the histograms of feature map values of 15th and 16th convolutional layers within SRResNet, respectively. For the super-resolution task, we empirically find that the data distribution ranges of the feature maps and weights within different layers are drastically different, as shown in (a) and (b) (mean 1.37 and std 2.23 for (a) and mean -2.88 and std 41.71 for (b)). Thus, we propose a trainable quantizer to adaptively decide the quantization interval according to the current distribution for mitigation of this phenomenon.}
\label{fig:daia-dist}
\end{figure}

In model quantization, the values within the quantization interval $I$ will be quantized. The quantization process would proceed smoothly if a suitable quantization interval is determined. However, once the quantization interval does not fit in the distribution of values to be quantized, it would incur a large quantization error. For the super-resolution task, we empirically find that the data distributions of the features and weights of different layers are drastically different, as shown in Figure \ref{fig:daia-dist}. Thus, different quantization intervals should be allocated for different quantizers. 
Toward this end, we propose to estimate the intervals automatically by parameterizing $I$.
To alleviate the optimization difficulty of the interval, we devise to find a good initial point for $I$ of a quantizer. Specifically, we propose to use the moving average of max values within the tensor $\bV_i$ (batch-wise activations or convolutional filters within a layer) to be quantized as the initial point:

\begin{equation}
    I = \frac{1}{l} \sum_{i=0}^{l-1} max(\bV_i).
\end{equation}

This process is performed at the first $l$ iterations of the model training as a warm-up. 
Then the parameterized interval $I$ is optimized in conjunction with other network parameters using backpropagation with stochastic gradient descent.
Similar to the training process of \cite{esser2019learned}, the gradient through the quantizer $Q(\cdot)$ to the quantization interval $I$ is approximated by STE as a pass-through function. 
Such that the intervals can be tuned in conjunction with other parameters of the model to further increase the representation ability of models.

\subsection{Fully Quantized Inference}

\begin{figure}[t]
\centering
\scalebox{0.52}{
\centerline{\includegraphics[width=1\textwidth]{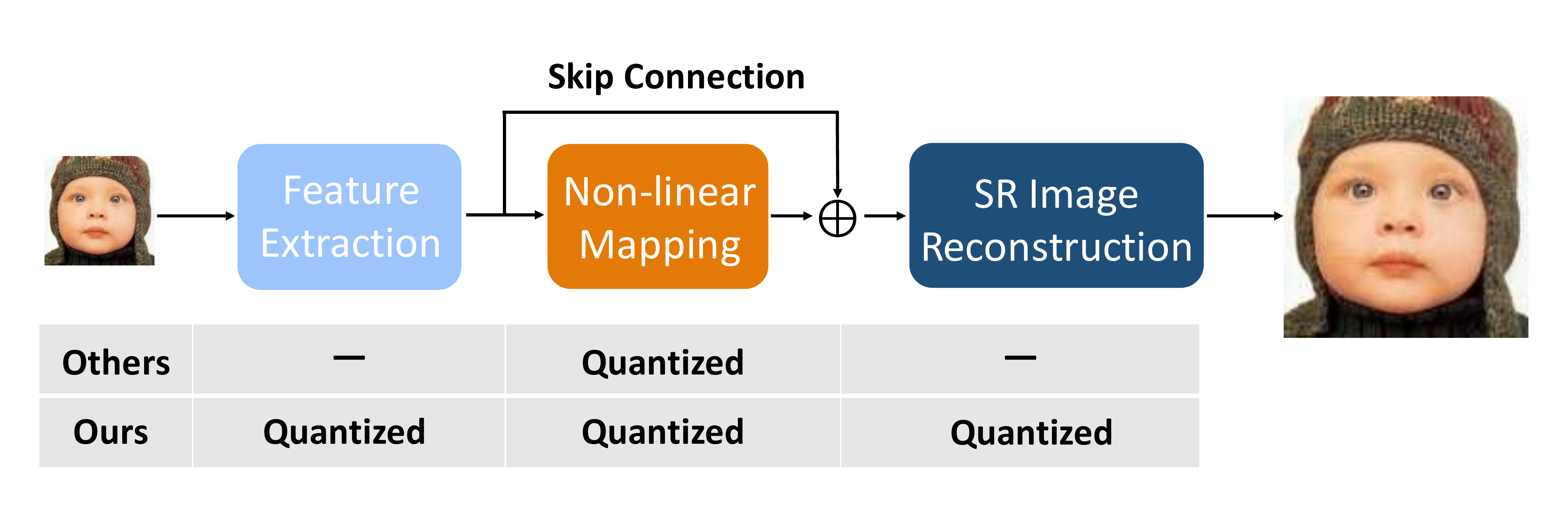}}
}
\caption{The overview of the proposed fully quantized super-resolution networks. The existing quantization models for super-resolution quantize the Non-linear Mapping part merely; while we quantize all three modules, by which large computation can be saved.}
\label{fig:framework}
\end{figure}

According to \cite{xin2020binarized}, the super-resolution process can be divided into three sub-modules: input feature extraction module $\mathcal{E}$, nonlinear mapping module $\mathcal{M}$ and SR image reconstruction module $\mathcal{R}$. 
Formally, for an input low-resolution image $lr$, the aforementioned process to generate a super-resolution image $sr$ can be presented as:

\begin{equation}
    sr=\mathcal{R}(\mathcal{M}(\mathcal{E}(lr))).
\end{equation}

Usually, there is a 
skip connection to link the feature extraction module $\mathcal{E}$ and the reconstruction module $\mathcal{R}$. Despite $\mathcal{E}$ and $\mathcal{R}$ consist of simple structures, they play an important role to achieve good performance in the super-resolution process. Moreover, large computational burdens are not only laid on the nonlinear mapping $\mathcal{M}$, but also on the reconstruction module $\mathcal{R}$, since the convolutional layers before upsampling are with a large number of channels to deal with the super-resolution output. However, current super-resolution quantization models propose to quantize $\mathcal{M}$ only \cite{ma2019efficient,xin2020binarized}. 
In addition, the existence of the skip connections within the network inevitably incurs huge memory consumption, which is known to dominate the energy consumption \cite{han2016deep}. 
To obtain an energy-efficient super-resolution framework, we propose to fully quantize all layers of the three modules, especially including all skip connections. The overview of the proposed fully quantized super-resolution network is shown in Figure \ref{fig:framework} and the comparison of quantization differences between existing SR quantization methods and our proposed FQSR networks is presented in Table \ref{tab:methods}.

\begin{table}[b]
\caption{Comparison of the quantized operations of different methods. Within the table, ``\checkmark'' represents whether quantization is enabled for the column; ``All modules'' include input feature extraction module $\mathcal{E}$, nonlinear mapping module $\mathcal{M}$ and SR image reconstruction module $\mathcal{R}$; ``wt'' stands for weight quantization of convolutional layers; ``fm'' denotes the feature map quantization; ``sc'' denotes the quantization of skip connections.}
\begin{center}
\resizebox{0.9\linewidth}{!}{
\begin{tabular}{l||c||c|c|c}
Methods      & All Modules & wt          & fm          & sc          \\ \hline
SRResNet\_Bin \cite{ma2019efficient} &            & \checkmark &            &            \\
SRGAN\_Bin \cite{ma2019efficient}    &            & \checkmark &            &            \\
VDSR\_BAM \cite{xin2020binarized}     &            & \checkmark & \checkmark &            \\
SRResNet\_BAM \cite{xin2020binarized} &            & \checkmark & \checkmark &            \\ \hline
FQSR (Ours)  & \checkmark & \checkmark & \checkmark & \checkmark
\end{tabular}
}\end{center}
\label{tab:methods}
\end{table}

\noindent\textbf{Quantization for BN.} 
During the inference phase, if the batch normalization layer is adopted in the quantized model, it can be folded into the preceding convolutional layer to get rid of the extra floating-point operations. The folding of the batch normalization operation is formally presented as:

\begin{equation}
\begin{aligned}
    z & =\gamma \left[\frac{(\bw \cdot \bx+b)-\mu}{\sqrt{\sigma^{2}+\epsilon}}\right]+\beta, \\
    & =\frac{\gamma \bw}{\sqrt{\sigma^{2}+\epsilon}} \cdot \bx+\frac{\gamma(b-\mu)}{\sqrt{\sigma^{2}+\epsilon}}+\beta, \\
    & =\bw_{fold} \cdot \bx + b_{fold},
\end{aligned}
\end{equation}
where $\bw$, $\bx$ and $b$ are the weights, inputs and bias term of the preceding convolutional layer, respectively; $\mu$ and $\sigma$ are the mean and standard deviation of the corresponding dimension; $z$ is the output of the batch normalization layer and $\bw_{fold}$, $b_{fold}$ are the weights and bias after folding, respectively.

\noindent\textbf{Quantization for skip connections.} Residual learning is critical to fetch exceptional representations in computer vision tasks. In the residual structural networks, skip connections are the core components to build direct links between shallow layers and deeper ones. Nevertheless, in quantized models, the skip connections carrying floating-point operands will hinder the model to be applied practically on embedding systems or mobile platforms because the quantization status of each layer is inconsistent. In addition, it will inevitably increase the computation as well. Moreover, for the super-resolution task, the input images and the images after super-resolution are usually in very high resolution (such as 2K or 4K). Therefore, the intermediate features conveyed through skip connections will occupy a huge amount of memory consumption.

In order to address the aforementioned issues, we quantize the skip connections through quantizing the output features of all convolutional layers and the element-wise addition layers. Consequently, the memory consumption will be saved dramatically (can be saved approximately $8\times$ when compared to the full-precision counterparts). Additionally, if the skip connections are quantized, the models are hardware-friendly since it is fully quantized. Formally, the element-wise addition operation of the skip connection in our quantized network can be formulated as:

\begin{equation}
    y = Q(x) + {\rm{ReLU}}(Q(z)).
\end{equation}

\begin{figure}[t]
\centering
\scalebox{0.48}{
\centerline{\includegraphics[width=1\textwidth]{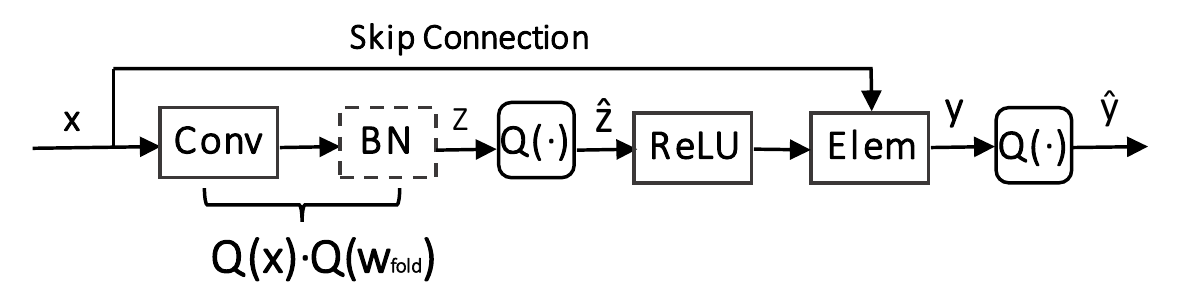}}
}
\caption{The process of the quantization in a typical residual block. In the process, $Q(\cdot)$ represents the quantization function. The input $x$ is the output of the preceding layer/residual block. The outputs of the convolutional layer $z$ and the element-wise layer $y$ are quantized to ensure the quantization of skip connections.}
\label{fig:ot-quant}
\end{figure}

Overall, the process of quantization in a typical residual block in the proposed FQSR network is shown in Figure \ref{fig:ot-quant}. Within the figure, $\hat{x}$ represents $Q(x)$.

\subsection{Self-supervised Quantization-aware Calibration Loss}
\label{sec:QCL}

\begin{figure}[h!]
\centering
\scalebox{0.56}{
\centerline{\includegraphics[width=1\textwidth]{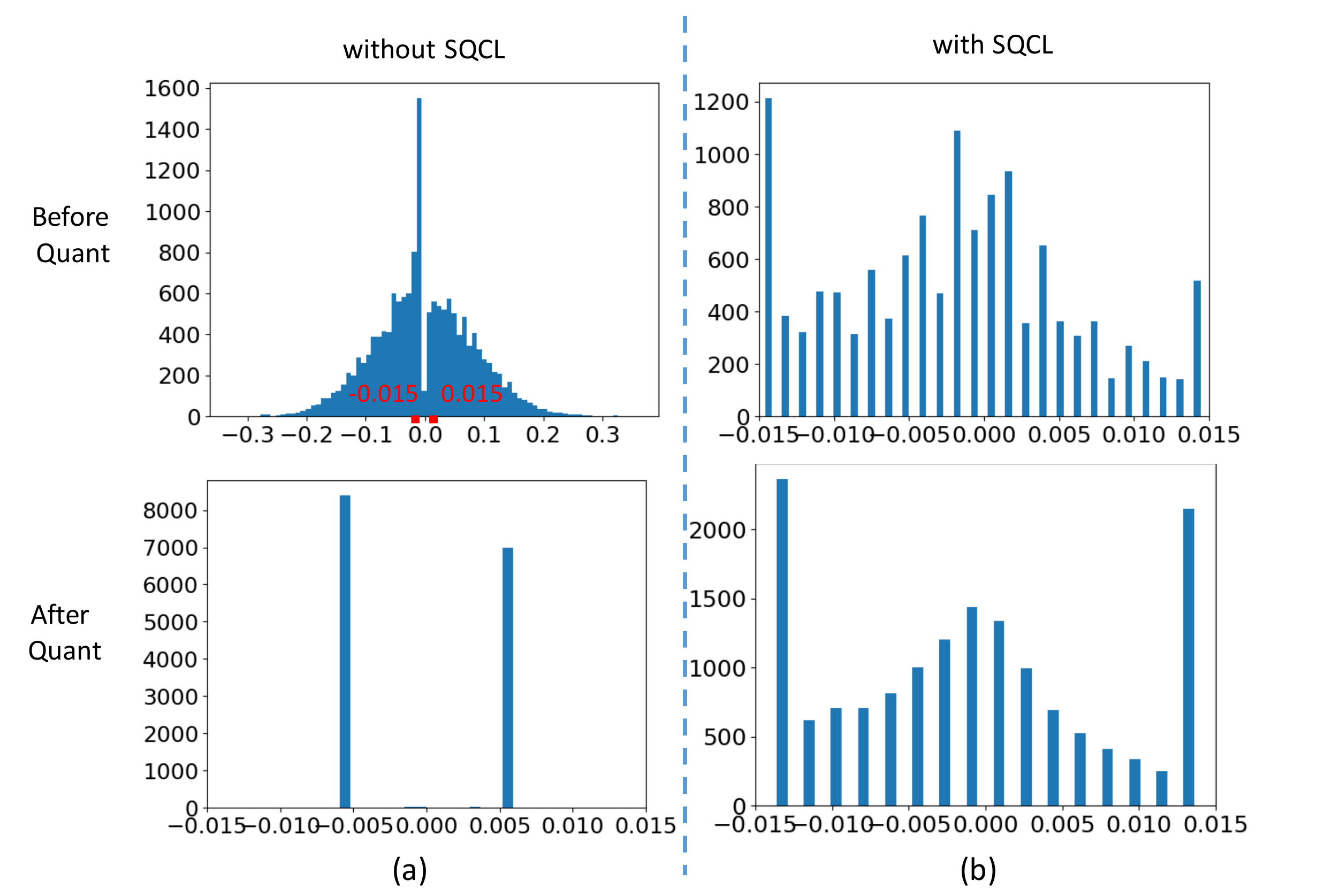}}
}
\caption{This figure shows, before and after adding the SQCL objective, the weights distribution of the last convolutional layer within SRResNet and its categorical distribution (4 bits, with 16 bins). Intuitively, the SQCL is adopted to minimize the jitters before and after quantization to constrain the model quantization in a smoother manner. Except for the sub-figure in the upper-left corner (since the range of the distribution within this sub-figure is hugely different from others), the other sub-figures are drawn under the range $[-0.015, 0.015]$. As shown in (a), for the model without SQCL objective, the data distributions before quantization and the categorical distribution after quantization are vastly different; With the SQCL objective as illustrated in (b), the categorical distribution is calibrated to fit the data distribution before quantization.}
\label{fig:qcl-dist}
\end{figure}

As shown in Figure \ref{fig:qcl-dist}, for the super-resolution task, we empirically observe that in some layers, especially in the last two layers, the data distributions before and after quantization change drastically. It is may be caused by the last two layers are responsible for upscaling purpose. Thus, the weight distributions are different from other feature transformation layers because of the dissimilar functionalities. However, this phenomenon will affect the model performance significantly due to the large quantization error. In order to minimize the quantization error, an objective function termed as Self-supervised Quantization-aware Calibration Loss (SQCL) is adopted to calibrate the values after quantization to have an approximate distribution as before quantization. In this case, the value before quantization serves as a strong self-supervision signal, which provides additional useful information for network optimization. The SQCL loss is applied to input activations, weights and outputs of each layer.

For a real value $\bv$ to be quantized, 
here we are targeting to find optimal parameters to minimize the difference of before and after quantization through back-propagation. Formally, SQCL is formulated as:

\begin{equation}
    \cL_q=\left\|Q(\bv)-\bv\right\|_{p},
    \label{eq:qcl}
\end{equation}
where $\parallel \cdot \parallel_p$ denotes the $L_p$ norm. 

Therefore the final objective function for the proposed super-resolution quantized networks is:

\begin{equation}
\label{eq:final}
    \cL = \cL_{sr} + \alpha \cL_q,
\end{equation}
where $L_{sr}$ represents the super-resolution loss and 
$\alpha$ is a balancing hyperparameter.

\noindent\textbf{Discussion.} 
The proposed quantization-aware calibration loss (SQCL) is different from the network distillation. In the design of network distillation, a teacher network (a larger or full precision model) is leveraged to train a student network (a small or low-bit quantization model). The teacher network is generally well-pretrained and provides ground truth to the student network. In contrast, our proposed quantization-aware calibration loss is an objective for self-supervised calibration of the quantizers, which encourages the quantized tensor to have a similar distribution to the original one. We employ the quantization-aware calibration loss for every quantization function, including weights and features, rather than only the network outputs.

Besides, the proposed quantization-aware calibration loss is also different from the previous quantization method LQ-net\cite{zhang2018lq}. In LQ-net, the authors propose to solve the non-uniform step sizes with a closed-form solution derived from a similar target in Eq. \ref{eq:qcl}. Unlike the LQ-net, we adopt the gradient-based loss function to stabilise the training process of the super-resolution quantization.

\section{Experiments}
\subsection{Experimental Setup}

Following the existing works \cite{lim2017enhanced,ledig2017photo,ma2019efficient,xin2020binarized}, we train our fully quantized super-resolution networks on DIV2K \cite{timofte2017ntire} dataset and evaluate models on five prevalent benchmark datasets. An extensive ablation study is further conducted to validate the effectiveness of each component within the proposed method.

\noindent\textbf{Datasets and evaluation metrics.} We conduct the model training on the DIV2k dataset, which is made up of 800 good quality high/low-resolution image pairs for model training, 100 image pairs for model validation and 100 image pairs for testing. However, the testing HR images for DIV2K is not publicly accessible, so we train models on 800 training images and validate models on 10 validation images. The best validation models are tested on Set5 \cite{bevilacqua2012low} (5 images), Set14 \cite{zeyde2010single} (14 images), BSD100 \cite{martin2001database} (100 images), Urban100 \cite{huang2015single} (100 images) and DIV2K (100 validation images). Two scaling settings are considered for model evaluation, containing $\times2$ and $\times4$.

For the super-resolution model evaluation, we take the most commonly adopted Peak Signal to Noise Ratio (PSNR) and Structural Similarity (SSIM) \cite{wang2004image} as our evaluation metrics. All evaluation is performed by cropping $s$ pixels for $\times s$ upscaling.

\noindent\textbf{Implementation details.} During training, random vertical/horizontal flips and 90-degree rotation are performed for data augmentation. The model batch size is set to 16 and Adam optimizer is adopted for model optimization. The initial learning rate is set to $1 \times 10^{-3}$ for SRResNet and SRGAN and $5 \times 10^{-5}$ for the EDSR model. The models are trained for 300 epochs with cosine annealing \cite{loshchilov2016sgdr} learning rate tuning strategy. The hyperparameter $l$ is set to 20 and the trade-off factor $\alpha$ is set to 0.3. $L_1$ norm is adopted for SQCL calculation. The models are implemented using PyTorch with one NVIDIA GTX 1080 Ti GPU. The experimental settings are fixed to all of our trained models to keep fair comparisons.

\subsection{Overall Performance}

We embed the proposed fully quantized super-resolution scheme on three state-of-the-art architectures, containing SRResNet, EDSR and SRGAN, to compare the effectiveness of low-bitwidth models with full-precision models and bicubic interpolation. The results are shown in Tables \ref{tab:compare}, \ref{tab:performance-edsr} and \ref{tab:performance-srgan}.

\noindent\textbf{Evaluation on SRResNet.} 
As shown in Table \ref{tab:compare}, we implement the FQSR on SRResNet with multiple configurations. Also, multiple state-of-the-art quantized SR models are compared with the proposed model. When compared to the bicubic interpolation, the 4/4/8 model (\ie, weights and activations are both quantized to 4 bits; skip connections are quantized to 8 bits) surpasses it by 2.678 with $\times 2$ up-scaling and 2.618 with $\times 4$ up-scaling for PSNR on Set5 dataset. 
By raising the skip connection precision to 8 bits, the performances are boosted by a large margin, \eg, 1.814 surpass on Set5 and 1.101 on Set14 for PSNR $\times 2$ up-scaling, respectively. It is worth noting that, on both $\times 2$ and $\times 4$ up-scaling settings, the 6/6/8 models achieve comparable results with or outperform the full-precision version of SRResNet. In addition, the 8/8/32 version fully quantized model significantly outperforms the full-precision counterpart %
by 0.121 and 0.336 on the PSNR metric with Set5 dataset for $\times 2$ and $\times 4$ up-scaling, respectively. The last rows of both $\times 2$ and $\times 4$ up-scaling settings are the lite version 6/6/8 configuration of the proposed FQSR model named as FQSR\_Lite, within which the nonlinear mapping $\mathcal{M}$ module only consists of 10 residual blocks rather than 16. Thus, we intend to save more computational cost while not losing much performance. On both $\times 2$ and $\times 4$ up-scaling settings, the FQSR\_Lite surpasses or receives comparable results with the 6/6/8 configuration but with a less computational cost.

\begin{table*}[htb]
\caption{The comparison between existing methods and our FQSR on SRResNet \cite{ledig2017photo}. The OPs are in the unit G Flops and Memory consumption is in the unit M Bytes. Similar to Table \ref{tab:methods}, ``wt'' represents weight quantization of convolutional layers; ``fm'' is the feature map quantization of layers; ``sc'' denotes the quantization of skip connections; ``p1'' represents the corresponding models are partially binarized.
}
\begin{center}
\resizebox{1.0\linewidth}{!}{
\begin{tabular}{l|llll|ll|llllllllll}
\toprule
Methods               & Scale & wt  & fm  & sc  & OPs   & Memo     & \multicolumn{2}{c}{Set5} & \multicolumn{2}{c}{Set14} & \multicolumn{2}{c}{B100} & \multicolumn{2}{c}{Urban100} & \multicolumn{2}{c}{DIV2K} \\
                      &       &    &    &    &         &          & PSNR        & SSIM       & PSNR         & SSIM       & PSNR        & SSIM       & PSNR          & SSIM         & PSNR         & SSIM       \\ \hline
SRResNet \cite{ledig2017photo}              & $\times 2$    & 32 & 32 & 32 & 997.018 & 531.117  & 37.760       & 0.958      & 33.270        & 0.914      & 31.950       & 0.895      & 31.280         & 0.919        & -            & -          \\
Bicubic               & $\times 2$    & 32 & 32 & 32 & -       & -        & 33.660       & 0.930       & 30.240        & 0.869      & 29.560       & 0.843      & 26.880         & 0.840         & 31.010        & 0.939      \\
SRResNet\_Bin \cite{ma2019efficient}  & $\times 2$    & p1 & 32 & 32 & 997.018 & 531.117  & 35.660       & 0.946      & 31.560        & 0.897      & -           & -          & 28.760         & 0.882        & -            & -          \\
SRResNet\_BAM \cite{xin2020binarized}          & $\times 2$    & p1 & p1 & 32 & 168.894 & 5842.287 & 37.210       & 0.956      & 32.740        & 0.910       & 31.600        & 0.891      & 30.200          & 0.906        & -            & -          \\
SRResNet\_DoReFa \cite{zhou2016dorefa}      & $\times 2$    & 8  & 8  & 8  & 124.627 & 132.779  & 37.205      & 0.956      & 32.967       & 0.911      & 31.837      & 0.894      & 30.974        & 0.916        & 31.050       & 0.939      \\
SRResNet\_w/o $\mathcal{M}$      & $\times 2$    & 32 & 32 & 32 & 155.749 & 177.039  & 36.863      & 0.954      & 32.536       & 0.907      & 31.379      & 0.887      & 29.525        & 0.896        & 33.268       & 0.934      \\ \hline
\multirow{5}{*}{FQSR (Ours)} & $\times 2$    & 4  & 4  & 8  & 62.314  & 132.779  & 36.338      & 0.945      & 32.403       & 0.901      & 31.367      & 0.882      & 29.982        & 0.899        & 32.357       & 0.927      \\
                      & $\times 2$    & 4  & 4  & 32 & 62.314  & 531.117  & 36.854      & 0.953      & 32.710        & 0.908      & 31.583      & 0.890       & 30.430         & 0.909        & 32.985       & 0.935      \\
                      & $\times 2$    & 6  & 6  & 8  & 93.470   & 132.779  & 37.541      & 0.957      & 33.236       & 0.913      & 31.966      & 0.894      & 31.398        & 0.920         & 33.964       & 0.941      \\
                      & $\times 2$    & 8  & 8  & 8  & 124.627 & 132.779  & 37.555      & 0.958      & 33.202       & 0.914      & 31.972      & 0.896      & 31.356        & 0.921        & 33.452       & 0.942      \\
                      & $\times 2$    & 8  & 8  & 32 & 124.627 & 531.117  & 37.881      & 0.959      & 33.408       & 0.915      & 32.093      & 0.897      & 31.712        & 0.924        & 34.424       & 0.943      \\
FQSR\_Lite (Ours)             & $\times 2$    & 6  & 6  & 8  & 63.894  & 132.779  & 37.349      & 0.956      & 33.070        & 0.912      & 31.851      & 0.895      & 30.964        & 0.916        & 33.753       & 0.939      \\ \hline \hline
SRResNet \cite{ledig2017photo}              & $\times 4$    & 32 & 32 & 32 & 383.487 & 132.777  & 31.760       & 0.888      & 28.250        & 0.773      & 27.380       & 0.727      & 25.540         & 0.767        & -            & -          \\
Bicubic               & $\times 4$    & 32 & 32 & 32 & -       & -        & 28.420       & 0.810       & 26.000           & 0.703      & 25.960       & 0.668      & 23.140         & 0.658        & 26.660        & 0.852      \\
SRResNet\_Bin \cite{ma2019efficient}  & $\times 4$    & p1 & 32 & 32 & 383.487 & 132.777  & 30.340       & 0.864      & 27.160        & 0.756      & -           & -          & 24.480         & 0.728        & -            & -          \\
SRResNet\_BAM \cite{xin2020binarized}          & $\times 4$    & p1 & p1 & 32 & 176.461 & 1460.580  & 31.240       & 0.878      & 27.970        & 0.765      & 27.150       & 0.719      & 24.950         & 0.745        & -            & -          \\
SRResNet\_DoReFa \cite{zhou2016dorefa}      & $\times 4$    & 8  & 8  & 8  & 47.936  & 33.194  & 31.539      & 0.885      & 28.156       & 0.771      & 27.299      & 0.724      & 25.384        & 0.763        & 28.132       & 0.825      \\
SRResNet\_w/o $\mathcal{M}$      & $\times 4$    & 32 & 32 & 32 & 173.175 & 44.260    & 30.880       & 0.841      & 27.808       & 0.723      & 27.059      & 0.694      & 24.777        & 0.714        & 28.081       & 0.815      \\ \hline
\multirow{5}{*}{FQSR (Ours)} & $\times 4$    & 4  & 4  & 8  & 23.968  & 33.194   & 31.038      & 0.874      & 27.860        & 0.761      & 27.090       & 0.714      & 24.949        & 0.744        & 27.925       & 0.816      \\
                      & $\times 4$    & 4  & 4  & 32 & 23.968  & 132.777  & 31.303      & 0.880       & 28.045       & 0.767      & 27.188      & 0.719      & 25.165        & 0.754        & 28.074       & 0.821      \\
                      & $\times 4$    & 6  & 6  & 8  & 35.952  & 33.194   & 31.923      & 0.889      & 28.404       & 0.775      & 27.452      & 0.727      & 25.752        & 0.774        & 28.571       & 0.830       \\
                      & $\times 4$    & 8  & 8  & 8  & 47.936  & 33.194   & 32.098      & 0.888      & 28.514       & 0.773      & 27.526      & 0.725      & 25.968        & 0.770         & 28.328       & 0.829      \\
                      & $\times 4$    & 8  & 8  & 32 & 47.936  & 132.777  & 32.096      & 0.892      & 28.559       & 0.780       & 27.555      & 0.732      & 26.034        & 0.783        & 28.894       & 0.836      \\
FQSR\_Lite (Ours)             & $\times 4$    & 6  & 6  & 8  & 28.558  & 33.194   & 31.644      & 0.886      & 28.249       & 0.774      & 27.348      & 0.727      & 25.460         & 0.765        & 28.405       & 0.826      \\ \hline
\end{tabular}
}\end{center}
\label{tab:compare}
\end{table*}

\noindent\textbf{Evaluation on EDSR.} In the evaluation on EDSR, as shown in Table \ref{tab:performance-edsr}, the 4/4/8 model is able to outperform bicubic interpolation by a large margin, 3.378 for $\times 2$ and 2.508 for $\times 4$ on Set5. On both $\times 2$ and $\times 4$ up-scaling settings, the 6/6/8 models achieve comparable results with the full-precision version of EDSR. The 8/8/8 and 8/8/32 models outperform the full-precision baseline model on most of the metrics. On $\times 2$, 0.137 PSNR improvement on Set14 and 0.283 PSNR improvement on Urban100 are obtained by the 8/8/32 model compared to the full-precision model.

\begin{table*}[htb]
\caption{The comparison of our FQSR with full-precision networks on EDSR \cite{lim2017enhanced} and Bicubic interpolation.}
\begin{center}
\resizebox{0.86\linewidth}{!}{
\begin{tabular}{l|llll|llllllllll}
\toprule
Methods               & Scale & wt  & fm  & sc  & \multicolumn{2}{c}{Set5} & \multicolumn{2}{c}{Set14} & \multicolumn{2}{c}{B100} & \multicolumn{2}{c}{Urban100} & \multicolumn{2}{c}{DIV2K} \\
                      &       &    &    &    & PSNR        & SSIM       & PSNR         & SSIM       & PSNR        & SSIM       & PSNR          & SSIM         & PSNR         & SSIM       \\ \hline
EDSR \cite{lim2017enhanced}  & $\times 2$ & 32 & 32 & 32 & 37.885       & 0.958      & 33.425       & 0.915      & 32.106      & 0.897      & 31.777        & 0.924        & 34.471       & 0.944      \\
Bicubic                      & $\times 2$ & 32 & 32 & 32 & 33.660       & 0.930      & 30.240       & 0.869      & 29.56       & 0.843      & 26.880        & 0.84         & 31.010       & 0.939      \\
EDSR\_DoReFa \cite{zhou2016dorefa}  & $\times 2$ & 8 & 8 & 8 & 37.849       & 0.958      & 33.418       & 0.915      & 32.096       & 0.897      & 31.746        & 0.924         & 34.374       & 0.943      \\ \hline
\multirow{5}{*}{FQSR (Ours)} & $\times 2$ & 4  & 4  & 8  & 37.038       & 0.951      & 32.835       & 0.908      & 31.668      & 0.889      & 30.646        & 0.911        & 33.282       & 0.933      \\
                             & $\times 2$ & 4  & 4  & 32 & 37.087       & 0.951      & 32.868       & 0.908      & 31.690      & 0.889      & 30.698        & 0.912        & 33.329       & 0.934      \\
                             & $\times 2$ & 6  & 6  & 8  & 37.817       & 0.958      & 33.411       & 0.915      & 32.092      & 0.897      & 31.784        & 0.925        & 34.306       & 0.943      \\
                             & $\times 2$ & 8  & 8  & 8  & 37.955       & 0.959      & 33.524       & 0.916      & 32.152      & 0.898      & 32.008        & 0.927        & 34.533       & 0.944      \\
                             & $\times 2$ & 8  & 8  & 32 & 37.993       & 0.959      & 33.562       & 0.916      & 32.171      & 0.898      & 32.060        & 0.927        & 34.586       & 0.945      \\ \hline \hline
EDSR \cite{lim2017enhanced}  & $\times 4$ & 32 & 32 & 32 & 32.007       & 0.892      & 28.486       & 0.778      & 27.528      & 0.731      & 25.934        & 0.781        & 28.880       & 0.835      \\
Bicubic                      & $\times 4$ & 32 & 32 & 32 & 28.420       & 0.810      & 26.000       & 0.703      & 25.960      & 0.668      & 23.140        & 0.658        & 26.660       & 0.852      \\
EDSR\_DoReFa \cite{zhou2016dorefa}  & $\times 4$ & 8 & 8 & 8 & 32.042       & 0.891      & 28.474       & 0.778      & 27.518       & 0.730      & 25.870        & 0.778         & 28.814       & 0.834      \\ \hline
\multirow{5}{*}{FQSR (Ours)} & $\times 4$ & 4  & 4  & 8  & 30.928       & 0.870      & 27.816       & 0.761      & 27.073      & 0.715      & 24.927        & 0.744        & 27.963       & 0.814      \\
                             & $\times 4$ & 4  & 4  & 32 & 30.983       & 0.872      & 27.856       & 0.762      & 27.090      & 0.716      & 24.947        & 0.746        & 27.976       & 0.815      \\
                             & $\times 4$ & 6  & 6  & 8  & 31.948       & 0.889      & 28.415       & 0.776      & 27.466      & 0.728      & 25.751        & 0.775        & 28.667       & 0.831      \\
                             & $\times 4$ & 8  & 8  & 8  & 32.102       & 0.892      & 28.524       & 0.779      & 27.539      & 0.731      & 25.944        & 0.781        & 28.853       & 0.835      \\
                             & $\times 4$ & 8  & 8  & 32 & 32.157       & 0.893      & 28.545       & 0.780      & 27.550      & 0.732      & 25.989        & 0.783        & 28.898       & 0.836      \\  \hline
\end{tabular}
}\end{center}
\label{tab:performance-edsr}
\end{table*}

\noindent\textbf{Evaluation on SRGAN.} The evaluation on SRGAN is shown in Table \ref{tab:performance-srgan}. Similar to the evaluation on SRResNet, the performance of 4/4/8 models achieve significantly better performance than bicubic interpolation on most of the metrics and datasets. Surprisingly, on $\times 2$ setting, the 6/6/8 setting outperforms the full-precision model on multiple metrics, \ie, 0.075 PSNR improvement on Set5 and 0.189 PSNR boost on Set14 compared with the full-precision model. Moreover, the 8/8/32 model outperforms the full-precision model on most of the metrics.

\begin{table*}[htb]
\caption{The comparison of Fully Quantized Super-resolution networks with full-precision networks on SRGAN \cite{ledig2017photo} and Bicubic interpolation.}
\begin{center}
\resizebox{0.86\linewidth}{!}{
\begin{tabular}{l|llll|llllllllll}
\toprule
Methods               & Scale & wt  & fm  & sc  & \multicolumn{2}{c}{Set5} & \multicolumn{2}{c}{Set14} & \multicolumn{2}{c}{B100} & \multicolumn{2}{c}{Urban100} & \multicolumn{2}{c}{DIV2K} \\
                      &       &    &    &    & PSNR        & SSIM       & PSNR         & SSIM       & PSNR        & SSIM       & PSNR          & SSIM         & PSNR         & SSIM       \\ \hline
SRGAN \cite{ledig2017photo}                 & $\times 2$    & 32 & 32 & 32 & 37.446      & 0.958      & 33.033       & 0.914      & 31.971      & 0.896      & 31.300          & 0.920         & 33.885       & 0.942      \\
Bicubic               & $\times 2$    & 32 & 32 & 32 & 33.660       & 0.930       & 30.240        & 0.869      & 29.560       & 0.843      & 26.880         & 0.840         & 31.010        & 0.939      \\
SRGAN\_DoReFa \cite{zhou2016dorefa}  & $\times 2$ & 8 & 8 & 8 & 37.344       & 0.956      & 33.045       & 0.912      & 31.876       & 0.894      & 31.133        & 0.918         & 31.431       & 0.939     \\ \hline
\multirow{5}{*}{FQSR (Ours)} & $\times 2$    & 4  & 4  & 8  & 36.693      & 0.950       & 32.644       & 0.906      & 31.565      & 0.888      & 30.373        & 0.908        & 32.921       & 0.933      \\
                      & $\times 2$    & 4  & 4  & 32 & 36.731      & 0.952      & 32.640        & 0.906      & 31.550       & 0.889      & 30.327        & 0.907        & 32.859       & 0.934      \\
                      & $\times 2$    & 6  & 6  & 8  & 37.521      & 0.957      & 33.222       & 0.913      & 31.955      & 0.894      & 31.343        & 0.919        & 33.975       & 0.941      \\
                      & $\times 2$    & 8  & 8  & 8  & 37.669       & 0.957      & 33.293       & 0.914      & 32.009        & 0.895      & 31.488        & 0.921        & 34.162        & 0.942      \\
                      & $\times 2$    & 8  & 8  & 32 & 37.665      & 0.958      & 33.254       & 0.914      & 31.980       & 0.895      & 31.378        & 0.919        & 34.060        & 0.941      \\ \hline \hline
SRGAN \cite{ledig2017photo}                 & $\times 4$    & 32 & 32 & 32 & 31.934      & 0.890       & 28.451       & 0.776      & 27.470       & 0.728      & 25.824        & 0.775        & 28.712       & 0.832      \\
Bicubic               & $\times 4$    & 32 & 32 & 32 & 28.420       & 0.810       & 26.000           & 0.703      & 25.960       & 0.668      & 23.140         & 0.658        & 26.660        & 0.852      \\
SRGAN\_DoReFa \cite{zhou2016dorefa}  & $\times 4$ & 8 & 8 & 8 & 31.351       & 0.883      & 28.074       & 0.770      & 27.254       & 0.723      & 25.294        & 0.760         & 28.184       & 0.824      \\ \hline
\multirow{5}{*}{FQSR (Ours)} & $\times 4$    & 4  & 4  & 8  & 30.963      & 0.872      & 27.854       & 0.759      & 27.078      & 0.713      & 24.932        & 0.742        & 27.833       & 0.814      \\
                      & $\times 4$    & 4  & 4  & 32 & 31.253      & 0.879      & 27.997       & 0.766      & 27.164      & 0.718      & 25.105        & 0.752        & 27.967       & 0.820       \\
                      & $\times 4$    & 6  & 6  & 8  & 31.874      & 0.889      & 28.398       & 0.775      & 27.443      & 0.726      & 25.732        & 0.772        & 28.625       & 0.830       \\
                      & $\times 4$    & 8  & 8  & 8  & 31.960       & 0.890       & 28.483       & 0.777      & 27.514      & 0.730       & 25.898        & 0.778        & 28.760        & 0.833      \\
                      & $\times 4$    & 8  & 8  & 32 & 32.030       & 0.891      & 28.482       & 0.778      & 27.499      & 0.729      & 25.864        & 0.777        & 28.793       & 0.833      \\ \hline
\end{tabular}
}\end{center}
\label{tab:performance-srgan}
\end{table*}

\subsection{Comparison with Existing SR Quantization Models}

The comparison of the proposed FQSR model with Ma \etal \cite{ma2019efficient} and Xin \etal \cite{xin2020binarized} on SRResNet is shown in Table \ref{tab:compare}, since they all provide results on SRResNet structure. Worth noting that, in \cite{ma2019efficient}, the models are trained 500 epochs for SRResNet and in \cite{xin2020binarized} the learning rate is decreased by half every 200 epoch, while we only train the FQSR model 300 epochs for comparison. With much fewer training epochs, the proposed FQSR models are able to achieve better performance with less computation cost and memory consumption. Following \cite{liu2020reactnet, wang2020differentiable,guo2020single}, OPs is the sum of low-bit operations and floating-point operations, \ie, for $M$-bit networks, OPs = BOPs/64 $\cdot M$ + FLOPs. Only the multiplication operations are calculated for OPs. In terms of memory consumption, because of the existence of long and short skip connections within the networks, we consider the peak memory consumption of each model at the inference stage. Maximally, feature maps of three convolutional layers are considered for SRResNet\_Bin and our proposed FQSR networks (one for long skip connection feature storing, one for short connection and another for the main trunk); the features of only one convolutional layer is considered for SRResNet\_w/o $\mathcal{M}$, since it just consists of three convolutional layers without skip connections. However, in terms of the SRResNet\_BAM model, because the activation quantization of each layer takes outputs of several preceding layers into consideration (these activations should be stored for re-using), features of 33 convolutional layers within $\mathcal{M}$ are computed. 
We consider 1020$\times$678 resolution DIV2K dataset images as inputs and $\times 2$ up-scaling as the configuration. The OPs are in the unit G (=$1 \times 10^{9}$) OPs and Memory consumption is in the unit M (=$1 \times 10^{6}$) Bytes.

In the table, SRResNet\_Bin is the binary SR network from paper \cite{ma2019efficient}. Because only the weights of each layer are quantized, floating-point operations are still required in the feature extraction module and image reconstruction module, as well as the skip connection to link these two modules. Thus, the OPs and memory consumption within in SRResNet\_Bin will not be reduced. SRResNet\_BAM is the bit accumulation model proposed by \cite{xin2020binarized}. It binarizes both the activation and weights of each convolutional layer, so it reduces the OPs and memory consumption to some extent. However, it does not take the quantization of convolutional layers before and after upsampling into consideration, which introduces huge OPs consumption. This is because in SR models, the convolutional channels should be raised before upsampling and the size of features is increased to their multiples after upsampling operations. What is more, the SRResNet\_BAM does not consider quantize the feature extraction module, image reconstruction module and the linking skip connection as well. SRResNet\_w/o $\mathcal{M}$ is the model that only consists of one convolutional layer within $\mathcal{E}$ and two convolutional layers within $\mathcal{R}$. The results show without $\mathcal{M}$, the simple full-precision super-resolution model could achieve comparable performance, such that the existence of a full-precision sub-net will shrink the significance of the model quantization dramatically. The results received by the simplified model represents the importance to quantize the Feature Extraction module and the
Reconstruction module.

If we compare SRResNet\_Bin with FQSR quantitatively, the 4/4/8 FQSR models (both $\times 2$ or $\times 4$ scales) are able to outperform SRResNet\_Bin models across almost all the metrics. When compared with SRResNet\_BAM algorithm, from the table, we can perceive that with approximately 1/2 of the OPs and 1/50 memory consumption only (FQSR 6/6/8 model) on $\times 2$ up-scaling, the FQSR model is able to achieve better results on multiple metrics and datasets (37.541 over 37.210 on PSNR for Set5). If we increase the bit number, the gaps will become bigger. Finally, our proposed lite version 6/6/8 model is able to receive better results compared to SRResNet\_BAM with remarkably fewer OPs. Furthermore, we compare FQSR with DoReFa \cite{zhou2016dorefa} models on the backbone of SRResNet, EDSR and SRGAN. As shown in table \ref{tab:performance-edsr}, for the 8/8/8 EDSR\_DoReFa model on both $\times 2$ and $\times 4$, the 8/8/8 FQSR model outperforms it across most of the metrics. Similar phenomenon is shown in table \ref{tab:performance-srgan} and table \ref{tab:compare} as well. This clearly shows the effectiveness of the proposed FQSR model.

\subsection{Ablation Study}

\begin{table*}[htb]
\caption{Ablation study on each component. The experiments are conducted on the 4/4/32 and $\times 2$ up-scaling setting.}
\begin{center}
\resizebox{0.8\linewidth}{!}{
\begin{tabular}{l|ll|llllllllll}
\toprule
Models & DAIA       & SQCL        & \multicolumn{2}{c}{Set5} & \multicolumn{2}{c}{Set14} & \multicolumn{2}{c}{B100} & \multicolumn{2}{c}{Urban100} & \multicolumn{2}{c}{DIV2K} \\
       &            &            & PSNR        & SSIM       & PSNR         & SSIM       & PSNR        & SSIM       & PSNR          & SSIM         & PSNR         & SSIM       \\ \hline
1      &            &            & 35.536      & 0.944      & 31.775       & 0.898      & 30.972      & 0.883      & 28.893        & 0.888        & 31.792       & 0.926      \\
2      & \checkmark &            & 36.372      & 0.945      & 32.395       & 0.901      & 31.397      & 0.884      & 30.164        & 0.902        & 32.455       & 0.927      \\
3      & \checkmark & \checkmark & 36.854      & 0.953      & 32.71        & 0.908      & 31.583      & 0.89       & 30.43         & 0.909        & 32.985       & 0.935      \\ \hline
\end{tabular}
}\end{center}
\label{tab:ablation}
\end{table*}

\begin{figure*}[htbp]
\centering
\subfigure{
\begin{minipage}[t]{0.39525\linewidth}
\centering
\includegraphics[width=1.05\textwidth]{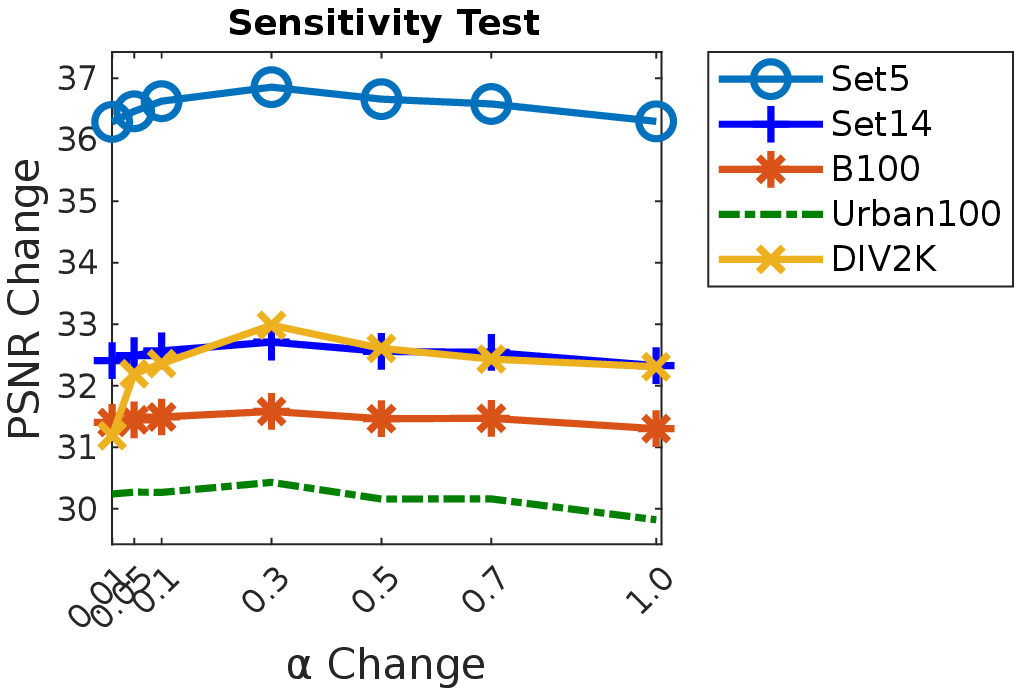}
\end{minipage}
}
\subfigure{
\begin{minipage}[t]{0.3952\linewidth}
\centering
\includegraphics[width=1.05\textwidth]{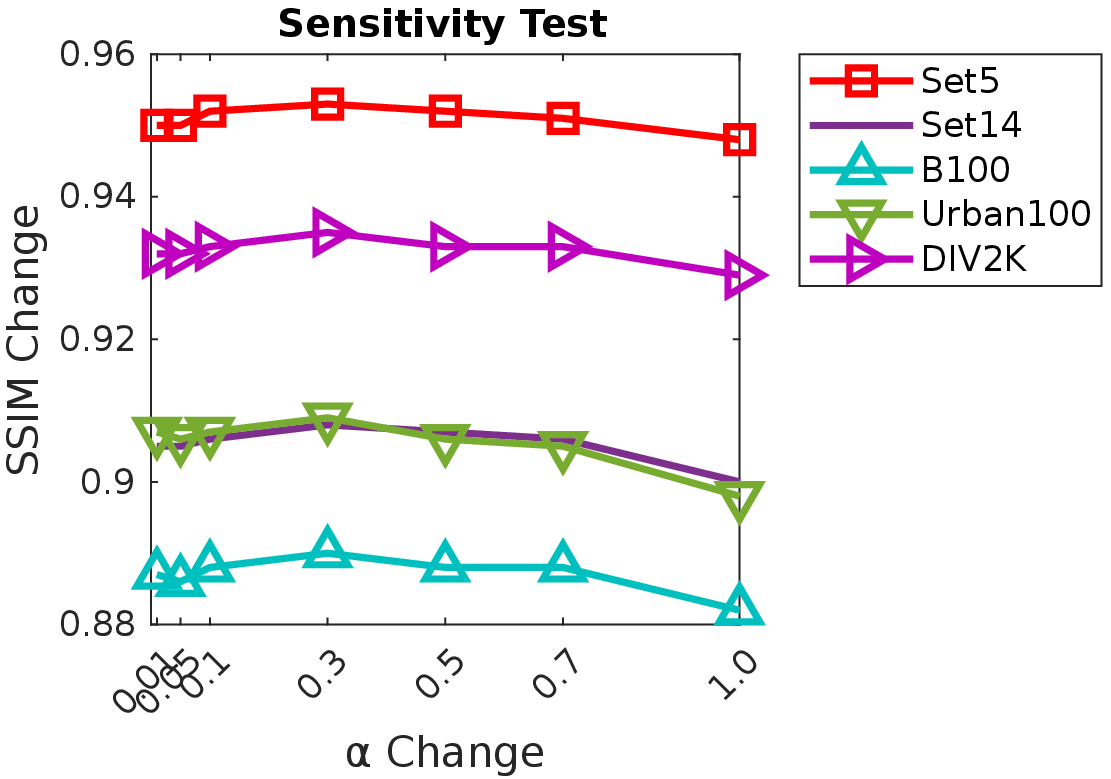}
\end{minipage}
}
\centering
\caption{PSNR and SSIM performance of FQSR using different loss trade-off factors on different datasets with $\times 2$ up-scaling.}\label{fig:sensitivity}
\end{figure*}

\noindent\textbf{Effect of different components.}
In this section, we examine the effect of each component in our FQSR model. 
The experimental results are reported in Table \ref{tab:ablation}. We empirically find that the proposed distribution-aware trainable quantization interval and the calibration loss are critical for the model to gain promising performance in the super-resolution process. With the trainable quantizers only, the PSNR performance of the baseline model is raised from 35.536 to 36.372 on Set5, and significant improvements on other metrics and datasets can also be observed. When equipped with strategies, the performance is further boosted to 36.854 on Set5. The ablation study generally shows the effectiveness of the proposed methods.

\noindent\textbf{Effect of $\alpha$.} This section presents the sensitivity of our FQSR model with different $\alpha$ setting in Eq. (\ref{eq:final}). Figure \ref{fig:sensitivity} demonstrates the PSNR and SSIM performance of FQSR on different datasets with the $\times 2$ up-scaling setting. It is clear that FQSR generally performs stably w.r.t. different $\alpha$ settings. 
From the figure, 
the trend of curves raising to peaks then falling can be observed for both PSNR and SSIM performance. In general, $\alpha = 0.3$ is recommended for the FQSR model to achieve the best performance.

\section{Conclusion}

In this paper, we have proposed a fully quantized super-resolution framework, including all layers within three SR sub-modules, as a practical solution to achieve a good trade-off between accuracy and efficiency. We have also identified difficulties faced by current low-bitwidth SR networks. That is 1) huge memory consumption caused by high-resolution feature maps 2) activation and weight distributions being vastly distinctive in different layers; 3) the inaccurate approximation of the quantization. In order to solve them, we quantize skip connections and two practical components have been proposed, a distribution-aware interval adaptation strategy to automatically decide the quantization intervals during training and a self-supervised quantization-aware calibration loss to explicitly minimize the quantization error. We have evaluated our method on multiple state-of-the-art deep super-resolution models on five benchmark datasets. The extensive experimental results have shown that our proposed FSQR is able to achieve state-of-the-art results while saving considerable computational cost and memory usage compared to the full-precision counterparts and competing methods. The ablation study further shows the proposed DAIA and SQCL are able to boost the model performance in a complementary manner.

\bibliographystyle{unsrt}
\bibliography{egbib}

\newpage
\appendix

\section{Visualization of Super-resolution Images}

\begin{figure*}[htb]
\centering
\scalebox{1.0}{
\centerline{\includegraphics[width=1\textwidth]{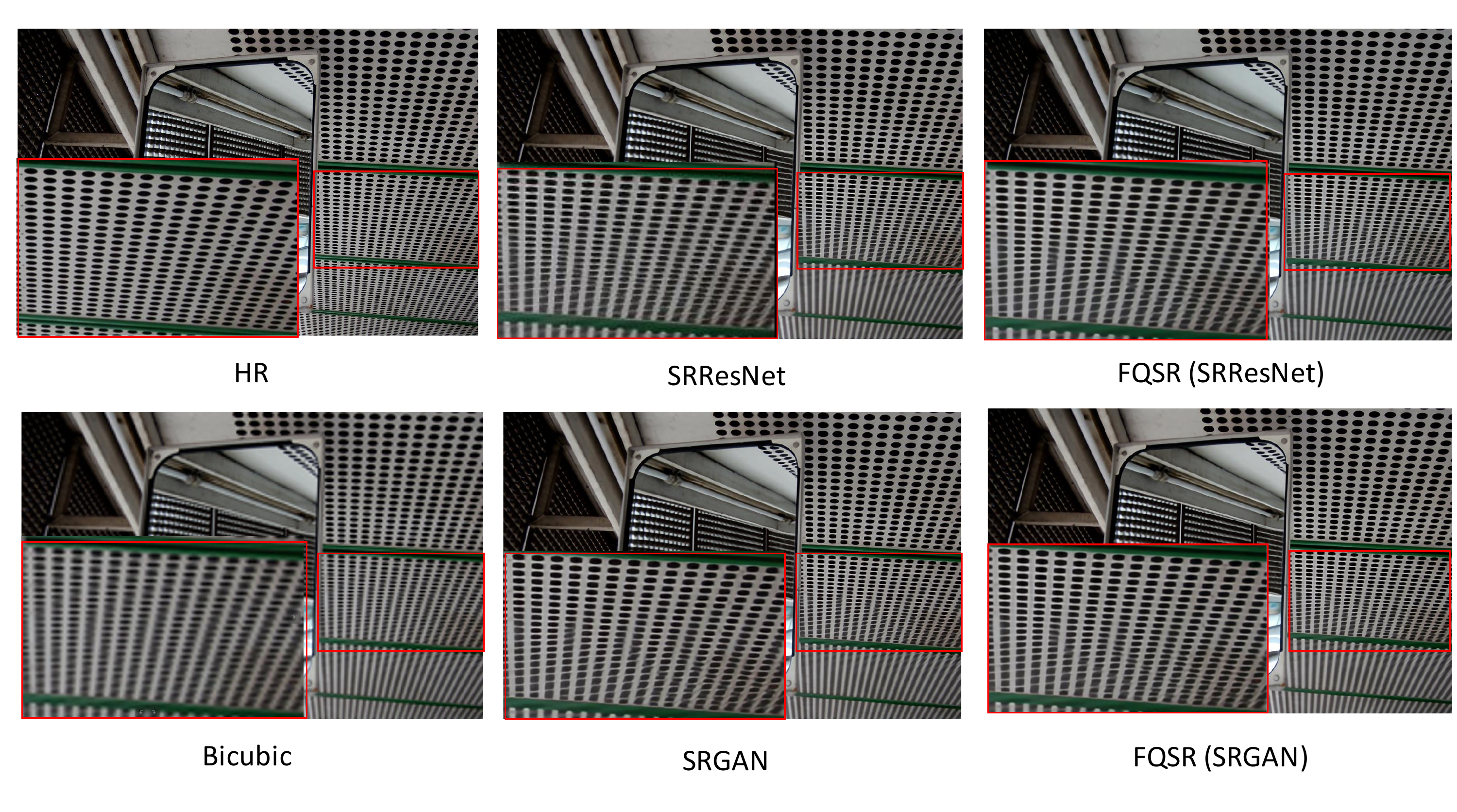}}
}
\caption{The visualization of Super-resolution images on $\times 4$ up-scaling. The SRResNet and SRGAN models denote full-precision models; the FQSR (SRResNet) and FQSR (SRGAN) models represent the 8/8/8 models on SRResNet and SRGAN respectively. As shown in the figure, the proposed FQSR models are able to achieve much better results than the bicubic method and comparable results with the full-precision models. The red box areas are the areas to be zoomed.}
\label{fig:sr-vis}
\end{figure*}

\begin{figure*}[htb]
\centering
\scalebox{1.0}{
\centerline{\includegraphics[width=1\textwidth]{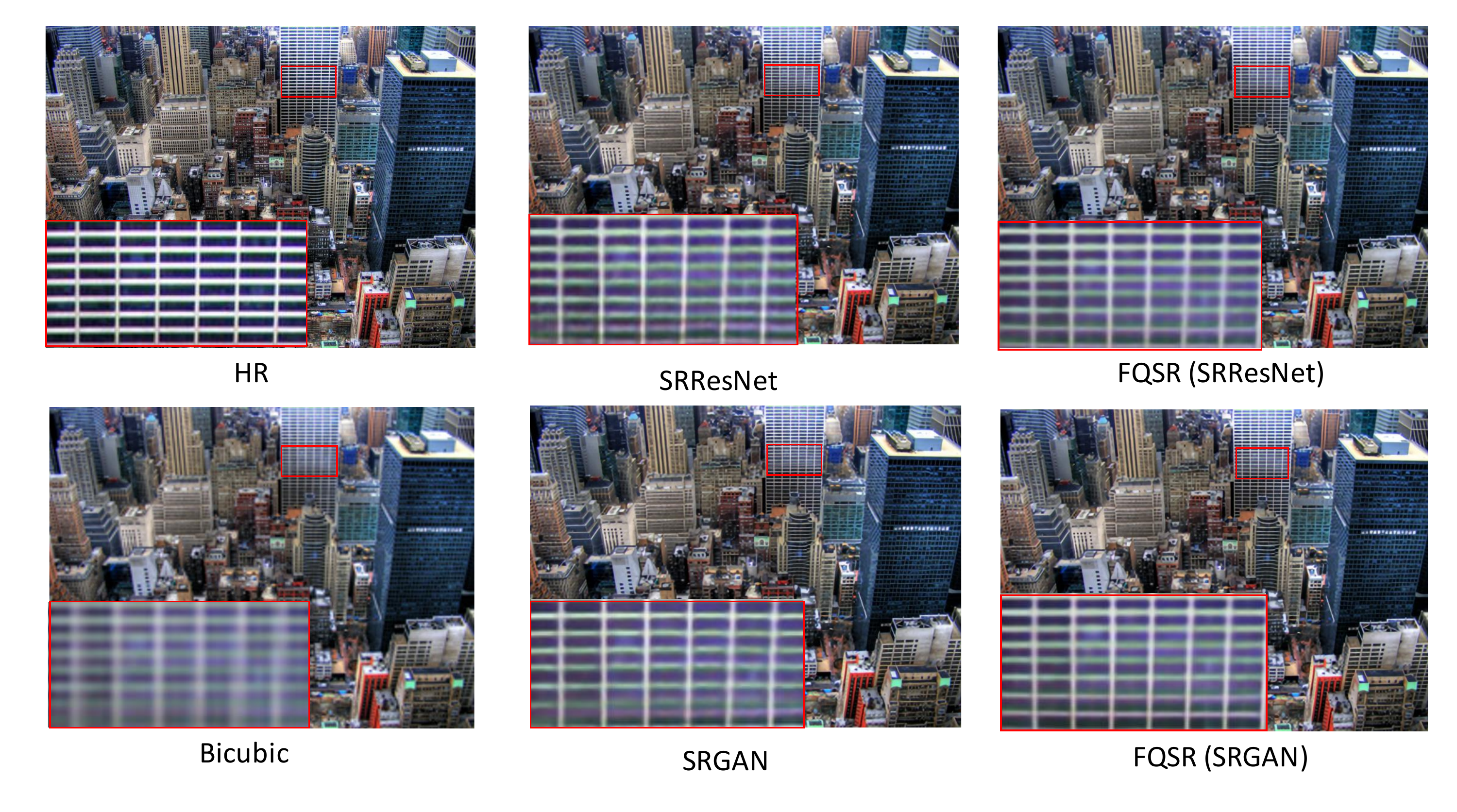}}
}
\caption{Another visualization of Super-resolution images on $\times 4$ up-scaling. As shown in the figure, the proposed FQSR models are able to achieve much better results than the bicubic method and comparable results with the full-precision models. The red box areas are the areas to be zoomed.}
\label{fig:sr-vis-2}
\end{figure*}

The visualization of super-resolution images are shown in Figure \ref{fig:sr-vis} and Figure \ref{fig:sr-vis-2}. The figure shows that the proposed FQSR models (8/8/8 model) are able to receive much better results than the bicubic method and comparable results with the full-precision models.

\section{Experimental Results with Self-ensemble}

Follow the evaluation of EDSR model \cite{lim2017enhanced}, in this section, we present the model performance with self-ensemble \cite{timofte2016seven}. The results generally show that with self-ensemble, the model performance can be further boosted.

\noindent\textbf{Self-ensemble \cite{timofte2016seven}} Self-ensemble is a strategy for SR model to further enhance the performance. During the testing phase, after rotating an input image at multiple angles, seven augmented images are obtained. By inputting the set of images into the model, the corresponding super-resolution images are obtained as well. Then, these super-resolution images are rotated back to the original angle. The final super-resolution image is obtained by weight-averaging these eight images (including the identity image).

\noindent\textbf{Evaluation on SRResNet \cite{ledig2017photo}} As shown in Table \ref{tab:performance-srresnet-ens}, when compared to the Bicubic interpolation, the performance of 4/4/8 model can surpass it by a large margin. 6/6/6, 6/6/8 models are able to receive comparable results with the full-precision models. 8/8/8 and 8/8/32 models can achieve better results than the full-precision models in most of the metrics and datasets. The lite version 6/6/8 model is able to outperform the normal 6/6/8 model with fewer computation requirements and memory consumption.

\begin{table*}[htb]
\caption{The comparison of our FQSR with full-precision networks on SRResNet with self-ensemble. The star signs represent models equipped with self-ensemble.}
\begin{center}
\resizebox{0.86\linewidth}{!}{
\begin{tabular}{l|llll|llllllllll}
\hline
Methods                        & Scale      & wt & fm & sc & \multicolumn{2}{c}{Set5} & \multicolumn{2}{c}{Set14} & \multicolumn{2}{c}{B100} & \multicolumn{2}{c}{Urban100} & \multicolumn{2}{c}{DIV2K} \\
                               &            &    &    &    & PSNR         & SSIM       & PSNR         & SSIM       & PSNR        & SSIM       & PSNR          & SSIM         & PSNR         & SSIM       \\ \hline
SRResNet* \cite{ledig2017photo} & $\times 2$ & 32 & 32 & 32 & 37.802       & 0.958      & 33.28        & 0.914      & 31.993      & 0.895      & 31.162        & 0.917        & 34.193       & 0.941      \\
Bicubic               & $\times 2$    & 32 & 32 & 32 & 33.660       & 0.930       & 30.240        & 0.869      & 29.56       & 0.843      & 26.880         & 0.84         & 31.010        & 0.939      \\ \hline
\multirow{7}{*}{FQSR* (Ours)}   & $\times 2$ & 4  & 4  & 8  & 37.099       & 0.955      & 32.752       & 0.91       & 31.646      & 0.891      & 30.253        & 0.908        & 33.018       & 0.937      \\
                               & $\times 2$ & 4  & 4  & 32 & 37.457       & 0.957      & 32.967       & 0.912      & 31.792      & 0.893      & 30.675        & 0.913        & 33.507       & 0.939      \\
                               & $\times 2$ & 6  & 6  & 6  & 37.703       & 0.957      & 33.285       & 0.913      & 32.021      & 0.895      & 31.413        & 0.92         & 34.019       & 0.941      \\
                               & $\times 2$ & 6  & 6  & 8  & 37.838       & 0.958      & 33.389       & 0.915      & 32.085      & 0.896      & 31.589        & 0.896        & 34.235       & 0.942      \\
                               & $\times 2$ & 8  & 8  & 8  & 37.796       & 0.959      & 33.328       & 0.916      & 32.049      & 0.897      & 31.556        & 0.923        & 33.634       & 0.943      \\
                               & $\times 2$ & 8  & 8  & 32 & 37.995       & 0.959      & 33.508       & 0.916      & 32.15       & 0.897      & 31.85         & 0.925        & 34.542       & 0.944      \\
FQSR\_Lite* (Ours)              & $\times 2$ & 6  & 6  & 8  & 37.673       & 0.958      & 33.249       & 0.913      & 31.97       & 0.895      & 31.16         & 0.918        & 34.051       & 0.941      \\ \hline \hline
SRResNet* \cite{ledig2017photo} & $\times 4$ & 32 & 32 & 32 & 32.106       & 0.892      & 28.567       & 0.778      & 27.552      & 0.73       & 25.919        & 0.777        & 28.891       & 0.834      \\
Bicubic               & $\times 4$    & 32 & 32 & 32 & 28.420       & 0.810       & 26.000           & 0.703      & 25.960       & 0.668      & 23.140         & 0.658        & 26.66        & 0.852      \\ \hline
\multirow{7}{*}{FQSR* (Ours)}   & $\times 4$ & 4  & 4  & 8  & 31.328       & 0.88       & 28.035       & 0.765      & 27.198      & 0.718      & 25.077        & 0.749        & 28.179       & 0.82       \\
                               & $\times 4$ & 4  & 4  & 32 & 31.613       & 0.885      & 28.213       & 0.771      & 27.299      & 0.722      & 25.294        & 0.758        & 28.306       & 0.825      \\
                               & $\times 4$ & 6  & 6  & 6  & 32.026       & 0.891      & 28.464       & 0.778      & 27.502      & 0.73       & 25.832        & 0.777        & 28.687       & 0.833      \\
                               & $\times 4$ & 6  & 6  & 8  & 32.102       & 0.892      & 28.509       & 0.778      & 27.523      & 0.729      & 25.884        & 0.777        & 28.729       & 0.833      \\
                               & $\times 4$ & 8  & 8  & 8  & 32.037       & 0.891      & 28.488       & 0.776      & 27.51       & 0.728      & 25.839        & 0.774        & 28.487       & 0.832      \\
                               & $\times 4$ & 8  & 8  & 32 & 32.25        & 0.894      & 28.633       & 0.782      & 27.611      & 0.733      & 26.149        & 0.786        & 28.998       & 0.837      \\
FQSR\_Lite* (Ours)              & $\times 4$ & 6  & 6  & 8  & 31.854       & 0.889      & 28.369       & 0.775      & 27.423      & 0.727      & 25.571        & 0.768        & 28.579       & 0.829      \\ \hline
\end{tabular}
}\end{center}
\label{tab:performance-srresnet-ens}
\end{table*}

\noindent\textbf{Evaluation on EDSR \cite{lim2017enhanced}} Table \ref{tab:performance-edsr-ens} shows the model performance on the EDSR structure. Similar as shown in Table \ref{tab:performance-srresnet-ens}, the 4/4/8 model is able to achieve much better performance than the Bicubic interpolation. The table generally similar results as the model without self-ensemble. The 6/6/6 version model can boost the performance significantly from 4/4/8 and 4/4/32. The 8/8/8 and 8/8/32 models can outperform the full-precision model in most cases.

\begin{table*}[htb]
\caption{The comparison of our FQSR with full-precision networks on EDSR with self-ensemble.}
\begin{center}
\resizebox{0.86\linewidth}{!}{
\begin{tabular}{l|llll|llllllllll}
\toprule
Methods               & Scale & wt  & fm  & sc  & \multicolumn{2}{c}{Set5} & \multicolumn{2}{c}{Set14} & \multicolumn{2}{c}{B100} & \multicolumn{2}{c}{Urban100} & \multicolumn{2}{c}{DIV2K} \\
                      &       &    &    &    & PSNR        & SSIM       & PSNR         & SSIM       & PSNR        & SSIM       & PSNR          & SSIM         & PSNR         & SSIM       \\ \hline
EDSR* \cite{lim2017enhanced}  & $\times 2$ & 32 & 32 & 32 & 38.014       & 0.959      & 33.564       & 0.916      & 32.188      & 0.898      & 32.009        & 0.926        & 34.620       & 0.944      \\
Bicubic                       & $\times 2$ & 32 & 32 & 32 & 33.660       & 0.930      & 30.240       & 0.869      & 29.560      & 0.843      & 26.880        & 0.840        & 31.010       & 0.939      \\ \hline
\multirow{7}{*}{FQSR* (Ours)} & $\times 2$ & 4  & 4  & 8  & 37.707       & 0.957      & 33.124       & 0.913      & 31.914      & 0.894      & 30.872        & 0.917        & 33.907       & 0.941      \\
                              & $\times 2$ & 4  & 4  & 32 & 37.726       & 0.958      & 33.161       & 0.913      & 31.938      & 0.895      & 30.922        & 0.917        & 33.962       & 0.941      \\
                              & $\times 2$ & 6  & 6  & 6  & 37.954       & 0.959      & 33.431       & 0.915      & 32.120      & 0.896      & 31.672        & 0.923        & 34.356       & 0.943      \\
                              & $\times 2$ & 6  & 6  & 8  & 38.044       & 0.959      & 33.527       & 0.916      & 32.181      & 0.898      & 31.929        & 0.926        & 34.541       & 0.945      \\
                              & $\times 2$ & 8  & 8  & 8  & 38.075       & 0.959      & 33.611       & 0.917      & 32.213      & 0.898      & 32.162        & 0.928        & 34.677       & 0.945      \\
                              & $\times 2$ & 8  & 8  & 32 & 38.084       & 0.959      & 33.645       & 0.917      & 32.221      & 0.898      & 32.207        & 0.929        & 34.703       & 0.945      \\ \hline \hline
EDSR* \cite{lim2017enhanced}  & $\times 4$ & 32 & 32 & 32 & 32.105       & 0.892      & 28.540       & 0.778      & 27.542      & 0.731      & 25.952        & 0.780        & 28.890       & 0.835      \\
Bicubic                       & $\times 4$ & 32 & 32 & 32 & 28.420       & 0.810      & 26.000       & 0.703      & 25.960      & 0.668      & 23.140        & 0.658        & 26.660       & 0.852      \\ \hline
\multirow{7}{*}{FQSR* (Ours)} & $\times 4$ & 4  & 4  & 8  & 31.263       & 0.880      & 27.998       & 0.768      & 27.218      & 0.722      & 25.042        & 0.752        & 28.310       & 0.824      \\
                              & $\times 4$ & 4  & 4  & 32 & 31.295       & 0.880      & 28.015       & 0.769      & 27.225      & 0.722      & 25.058        & 0.753        & 28.304       & 0.825      \\
                              & $\times 4$ & 6  & 6  & 6  & 31.956       & 0.890      & 28.427       & 0.777      & 27.473      & 0.729      & 25.718        & 0.774        & 28.699       & 0.832      \\
                              & $\times 4$ & 6  & 6  & 8  & 32.121       & 0.892      & 28.529       & 0.779      & 27.542      & 0.731      & 25.855        & 0.779        & 28.849       & 0.835      \\
                              & $\times 4$ & 8  & 8  & 8  & 32.250       & 0.894      & 28.632       & 0.781      & 27.604      & 0.733      & 26.078        & 0.785        & 28.979       & 0.837      \\
                              & $\times 4$ & 8  & 8  & 32 & 32.275       & 0.895      & 28.652       & 0.782      & 27.612      & 0.733      & 26.124        & 0.786        & 29.014       & 0.838      \\ \hline
\end{tabular}
}\end{center}
\label{tab:performance-edsr-ens}
\end{table*}

\noindent\textbf{Evaluation on SRGAN \cite{ledig2017photo}} Table \ref{tab:performance-srgan-ens} shows the model performance on the SRGAN structure. When compared to Bicubic interpolation, the performance of 4/4/8 model can outperform it significantly. Surprisingly, the 6/6/8 models surpass or receive comparable results to the full-precision models in multiple datasets and metrics.

\begin{table*}[htb]
\caption{The comparison of our FQSR with full-precision networks on SRGAN with self-ensemble.}
\begin{center}
\resizebox{0.86\linewidth}{!}{
\begin{tabular}{l|llll|llllllllll}
\toprule
Methods               & Scale & wt  & fm  & sc  & \multicolumn{2}{c}{Set5} & \multicolumn{2}{c}{Set14} & \multicolumn{2}{c}{B100} & \multicolumn{2}{c}{Urban100} & \multicolumn{2}{c}{DIV2K} \\
                      &       &    &    &    & PSNR        & SSIM       & PSNR         & SSIM       & PSNR        & SSIM       & PSNR          & SSIM         & PSNR         & SSIM       \\ \hline
SRGAN* \cite{ledig2017photo}  & $\times 2$ & 32 & 32 & 32 & 37.741       & 0.958      & 33.357       & 0.915      & 32.033      & 0.897      & 31.611        & 0.923        & 33.988       & 0.943      \\
Bicubic               & $\times 2$    & 32 & 32 & 32 & 33.660       & 0.930       & 30.240        & 0.869      & 29.56       & 0.843      & 26.880         & 0.84         & 31.010        & 0.939      \\ \hline
\multirow{7}{*}{FQSR* (Ours)} & $\times 2$ & 4  & 4  & 8  & 37.46        & 0.957      & 32.979       & 0.911      & 31.824      & 0.893      & 30.698        & 0.913        & 33.495       & 0.938      \\
                             & $\times 2$ & 4  & 4  & 32 & 37.454       & 0.957      & 32.972       & 0.911      & 31.792      & 0.893      & 30.61         & 0.912        & 33.372       & 0.938      \\
                             & $\times 2$ & 6  & 6  & 6  & 37.712       & 0.957      & 33.302       & 0.913      & 32.015      & 0.895      & 31.416        & 0.92         & 33.994       & 0.941      \\
                             & $\times 2$ & 6  & 6  & 8  & 37.809       & 0.958      & 33.374       & 0.915      & 32.066      & 0.896      & 31.554        & 0.922        & 34.236       & 0.942      \\
                             & $\times 2$ & 8  & 8  & 8  & 37.897       & 0.959      & 33.43        & 0.915      & 32.106      & 0.897      & 31.689        & 0.923        & 34.378       & 0.943      \\
                             & $\times 2$ & 8  & 8  & 32 & 37.851       & 0.959      & 33.351       & 0.915      & 32.048      & 0.896      & 31.442        & 0.921        & 34.229       & 0.943      \\ \hline \hline
SRGAN* \cite{ledig2017photo}  & $\times 4$ & 32 & 32 & 32 & 32.079       & 0.891      & 28.548       & 0.778      & 27.532      & 0.729      & 25.936        & 0.778        & 28.815       & 0.834      \\
Bicubic               & $\times 4$    & 32 & 32 & 32 & 28.420       & 0.810       & 26.000           & 0.703      & 25.960       & 0.668      & 23.140         & 0.658        & 26.66        & 0.852      \\ \hline
\multirow{7}{*}{FQSR* (Ours)} & $\times 4$ & 4  & 4  & 8  & 31.291       & 0.879      & 27.998       & 0.764      & 27.185      & 0.717      & 25.042        & 0.747        & 28.109       & 0.82       \\
                             & $\times 4$ & 4  & 4  & 32 & 31.56        & 0.885      & 28.183       & 0.77       & 27.285      & 0.722      & 25.248        & 0.756        & 28.238       & 0.824      \\
                             & $\times 4$ & 6  & 6  & 6  & 32.008       & 0.89       & 28.455       & 0.777      & 27.488      & 0.729      & 25.808        & 0.775        & 28.464       & 0.831      \\
                             & $\times 4$ & 6  & 6  & 8  & 32.074       & 0.891      & 28.5         & 0.777      & 27.517      & 0.728      & 25.858        & 0.776        & 28.791       & 0.832      \\
                             & $\times 4$ & 8  & 8  & 8  & 32.163       & 0.893      & 28.576       & 0.779      & 27.568      & 0.731      & 26.003        & 0.781        & 28.872       & 0.835      \\
                             & $\times 4$ & 8  & 8  & 32 & 32.173       & 0.893      & 28.571       & 0.779      & 27.562      & 0.731      & 25.971        & 0.78         & 28.9         & 0.835      \\ \hline
\end{tabular}
}\end{center}
\label{tab:performance-srgan-ens}
\end{table*}

\end{document}